\documentclass[aps,preprint,groupedaddress,showpacs,nofootinbib,amsmath,
amssymb]
{revtex4}

\usepackage{bm,feynmf,bbm,feynmf}

\begin{document}

\title{$\bm{SU(4)_C \times SU(2)_L \times SU(2)_R}$ Models With
$\bm{C}$-parity}

\author{Chin-Aik Lee}
% \email{jlca@udel.edu}
\affiliation{Bartol Research Institute, Department of Physics \& Astronomy,
University of Delaware,
Newark, DE 19716}

\author{Qaisar Shafi}
% \email{shafi@bartol.udel.edu}
\affiliation{Bartol Research Institute, Department of Physics \& Astronomy,
University of Delaware,
Newark, DE 19716}

\date{\today}

\begin{abstract}
We construct supersymmetric $SU(4)_C \times SU(2)_L \times SU(2)_R$ models
with
spontaneously broken left-right symmetry ($C$-parity). The minimal 
supersymmetric standard model (MSSM) can be recovered at low scales by
exploiting the missing partner mechanism. The field content is 
compatible with realistic fermion masses and mixings, proton
lifetime is close to or exceeds the current experimental bounds, and 
supersymmetric hybrid inflation can be implemented to take care of 
$C$-parity domain walls as well as magnetic monopoles, and to realize the 
observed baryon asymmetry via non-thermal leptogenesis.
\end{abstract}

\preprint{BA-07-35}

\pacs{11.15.Ex, 11.30.Er, 12.60.Jv, 14.80.Cp}

\maketitle

In a previous article \cite{Lee:2007aa} we constructed supersymmetric $SU(3)_C
\times 
SU(2)_L \times SU(2)_R \times U(1)_{B-L}$ models \cite{Pati:1974yy,lr} in which
the scalar (Higgs) sector respects 
a spontaneously broken discrete left-right symmetry ($C$-parity)
\cite{Lazarides:1985my,Kibble:1982ae}. A 
variety of symmetry breaking scales were discussed, and it was shown 
that for TeV scale breaking, a large number of new particles potentially 
much lighter than the $SU(2)_R$ charged gauge boson could be found at the 
LHC. This current article is a continuation of \cite{Lee:2007aa} and here we
will explore 
supersymmetric models based on the well known gauge symmetry $G_{422} \equiv 
SU(4)_C \times SU(2)_L \times SU(2)_R$ \cite{Pati:1974yy}.
Being a maximal subgroup of $Spin(10)$ (also known as $SO(10)$), $G_{422}$ 
captures many of its most salient features. For instance, $G_{422}$ gives 
rise to electric charge quantization, explains the standard model 
quantum numbers of each family, and predicts the existence of right 
handed neutrinos. However, there are also some important differences
between $SO(10)$ and $G_{422}$ which can be experimentally tested. For
instance, in $G_{422}$ the lightest magnetic monopole carries two quanta 
of Dirac magnetic charge \cite{Lazarides:1980cc}. (In $SO(10)$ the lightest
monopole carries 
one quantum of Dirac magnetic charge, unless $SO(10)$ breaks via $G_{422}$.) 
By the same token in the absence of $SO(10)$, $G_{422}$ predicts the existence 
of $SU(3)$ color singlet states carrying electric charges $\pm e/2$
\cite{Lazarides:1980cc,King:1997ia,Kephart:2006zd}. 
Finally, gauge coupling unification and gauge boson mediated proton 
decay are a characteristic feature of $SO(10)$
($C$-parity reduces from three to two the number of independent gauge couplings
in
$G_{422}$).
Following \cite{Lee:2007aa}, we construct $G_{422}$ based supersymmetric
models 
supplemented by $C$-parity. Ref. \cite{Melfo:2003xi} considered similar models
but there 
are some important differences. For instance, we exploit a missing 
partner mechanism \cite{mpSU5} to realize MSSM at low energies
without fine tuning. In our models, among other things, we also can 
realize supersymmetric inflation to avoid the monopole problem and 
$C$-parity domain walls.

In the simplest $G_{422}$ models, the MSSM electroweak doublets come from a
bidoublet
$H(\bm{1},\bm{2},\bm{2})$, the matter fields are unified into three generations
of $\Psi(\bm{4},\bm{2},\bm{1})$, the antimatter fields into three generations
of $\Psi^c(\bm{\overline{4}},\bm{1},\bm{2})$, and the Yukawa couplings for
matter come from $H\Psi^c\Psi$. In the simplest such mechanism, we are left
with the unwanted relation $Y^U=Y^D=Y^E=Y^{Dirac}$ between the Yukawa couplings,
which can at best match experimental data for the third generation. We will
discuss
later how to get around this.

Two of the simplest ways of breaking $G_{422}$ down to MSSM is to either use
$(\bm{4},\bm{1},\bm{2})$/$(\bm{\overline{4}},\bm{1},\bm{2})$, which we will
call $\Phi_R^c$ and $\Phi_R$ respectively and/or
$(\bm{10},\bm{1},\bm{3})$/$(\bm{\overline{10}},\bm{1},\bm{3})$ superfields
which we will call $\Delta_R^c$ and $\Delta_R$ respectively. (This is
analogous to the $\Phi$'s and $\Delta$'s of Ref. \cite{Lee:2007aa}, except
that these fields also contain color triplets). By invoking
$C$-parity, we also ensure that these chiral superfields will come with their
$C$-conjugates, namely $\Phi_L(\bm{4},\bm{2},\bm{1})$,
$\Phi_L^c(\bm{\overline{4}},\bm{2},\bm{1})$, $\Delta_L
(\bm{10},\bm{3},\bm{1})$ and $\Delta_L^c (\bm{\overline{10}},\bm{3},\bm{1})$.

We begin by analyzing a model with only $\Phi$'s and no $\Delta$'s.
Unlike the $SU(2)_L \times SU(2)_R \times U(1)_{B-L}$ models we considered in
\cite{Lee:2007aa}, the $\Phi$'s here
decompose into $SU(3)_C$ singlets as well as triplets. The mechanisms presented
there can serve to pair up the color singlet components but will fail to pair
up the color triplet components. Let us see why this is the case. We
double the number of bidoublets and consider the following terms in the
superpotential W:
\begin{equation}
W \supset \kappa S(\Phi_L^c \Phi_L + \Phi_R^c \Phi_R - M^2), H_1 \Phi_L \Phi_R,
H_2
\Phi_L^c \Phi_R^c,
\end{equation}
where $S$ is a gauge singlet superfield and $H_1$, $H_2$ are the two bidoublets.
Following \cite{Lee:2007aa}, we will find that we get a solution with nonzero
VEVs for the $\Phi_R$'s but not the $\Phi_L$'s and that as a result of this, the
up-type Higgs component of the bidoublet $H_1$ pairs up with the down-type Higgs
component of $\Phi_L$, while the down-type Higgs component of $H_2$ pairs up
with
the up-type Higgs component of $\Phi_L^c$. This is the so-called missing
partner mechanism \cite{mpSU5} because what remains at low energies of the
bidoublets is the
down-type
component of $H_1$ and the up-type component of $H_2$. Now, the up-type Yukawa
couplings can come from $H_2 \Psi^c \Psi$ and the down-type
couplings from $H_1 \Psi^c \Psi$. Because of this, we no longer have
any relation between $Y^U$ and $Y^D$. 

However, in $G_{422}$, the $\Phi_L$'s also
contain
$(\bm{3},\bm{2})_{\frac{1}{3}}$ and $(\bm{\overline{3}},\bm{2})_{-\frac{1}{3}}$
components (in MSSM notation) and those still remain unpaired. Similarly, some
linear combinations of the color singlet components of the $\Phi_R$'s pair up
with $S$ and the others become Goldstone and sgoldstones. The color triplets
$(\bm{3},\bm{1})_{\frac{4}{3}}$ and $(\bm{\overline{3}},\bm{1})_{-\frac{4}{3}}$
also become Goldstones and sgoldstones but the other color triplets
$(\bm{3},\bm{1})_{-\frac{2}{3}}$ and $(\bm{\overline{3}},\bm{1})_{\frac{2}{3}}$
do not. 

A solution to this proposed in
\cite{King:1997ia,Antoniadis:1988cm} is to introduce a
$(\bm{6},\bm{1},\bm{1})$ Higgs field
and the couplings $(\bm{6},\bm{1},\bm{1})\Phi_R\Phi_R$ and
$(\bm{6},\bm{1},\bm{1})\Phi_R^c\Phi_R^c$. Here, however, we also have $\Phi_L$
and $\Phi_L^c$ Higgs fields containing $(\bm{3},\bm{2})_{\frac{1}{3}}$ and
$(\bm{\overline{3}},\bm{2})_{-\frac{1}{3}}$ components. Thus, instead of
$(\bm{6},\bm{1},\bm{1})$, we introduce a $(\bm{15},\bm{1},\bm{1})$
Higgs superfield and add the following terms to W:
\begin{align}
W \supset \alpha H_{15} (\Phi_L \Phi_L^c + \Phi_R \Phi_R^c) + M'H_{15}^2.
\end{align}
This induces a nonzero VEV along the MSSM singlet direction for $H_{15}$. 

By
varying with
respect to $\Phi_R$ and $\Phi_R^c$, we find that the $44$-component of $\kappa S
\mathbbm{1}_{4\times 4} + \alpha H_{15}$ has to be zero. This means that the
color triplet components of $\Phi_L$ and $\Phi_R$ get paired up but the color
singlet components do not. The non-zero VEVs are as follows:
\begin{subequations}
\begin{align}
\langle \Phi_R \rangle = \langle \Phi_R^c \rangle &=
\begin{pmatrix}0&0\\0&0\\0&0\\M&0\end{pmatrix}\\
\langle H_{15} \rangle &=
\alpha\frac{M^2}{M'}\begin{pmatrix}\frac{1}{8}&0&0&0\\0&\frac{1}{8}
&0&0\\0&0&\frac{1}{8}&0\\0&0&0&-\frac{3}{8}\end{pmatrix}
\label{eq:15VEV}\\
\langle S \rangle &= \frac{3}{8}\frac{\alpha^2}{\kappa}\frac{M^2}{M'}\\
\kappa S \mathbbm{1} + \alpha H_{15} &=
\alpha^2\frac{M^2}{M'}\begin{pmatrix}\frac{1}{2}&0&0&0\\0&\frac{1}{2}
&0&0\\0&0&\frac{1}{2}&0\\0&0&0&0\end{pmatrix}
\label{eq:PhiR_mass}
\end{align}
\end{subequations}
The $(\bm{1},\bm{1})_{\pm 2}$ components become the
Goldstone and the sgoldstones. $S$ pairs up with a linear combination of
$(\bm{1},\bm{1})_0$ and the orthogonal combination form the goldstone
multiplet. The $2\times 2$ mass matrix for the $(\bm{3},\bm{1})_{\frac{4}{3}}$
and
$(\bm{\overline{3}},\bm{1})_{-\frac{4}{3}}$ components in both $\Phi_R$ and
$H_{15}$ is given by
\begin{equation}
\begin{pmatrix}
\frac{1}{2}\alpha^2 \frac{M^2}{M'} & \alpha M\\
\alpha M & 2M'
\end{pmatrix}
\label{eq:triplet_mass}
\end{equation}
which has a zero determinant. The direction with the zero eigenvalue is both the
goldstone and sgoldstone direction. 

Note that the unpaired set
of weak doublets from $\Phi_L$ and $\Phi_L^c$ have a mass term which
goes as $\kappa \langle S \rangle \mathbbm{1} + \alpha H_{15}$. This is
where the missing partner mechanism involving two Higgs bidoublets comes in
handy, as explained earlier. Because of the $SU(4)_C$ symmetry, the
model as it stands still
suffers from the unwanted relation $Y^D=Y^E$. Now it turns out that there are a
number of ways around this problem but we will only deal with the two most
common strategies here; one of them is to introduce the nonrenormalizable
coupling $H_1 H_{15} \Psi^c \Psi/\Lambda$ and the other is to introduce a
$(\bm{15},\bm{2},\bm{2})$ Higgs field \cite{Lazarides:1980nt}. Eq.
\ref{eq:15VEV} tells us that the
contribution of $H_1 H_{15} \Psi^c \Psi/\Lambda$ to $Y^D$ is
$-\frac{1}{3}$ that of the
contribution to $Y^E$. If $\Lambda$ corresponds to some value an order of
magnitude or so larger than the GUT scale,
and if the $H_1 \Psi^c_3 \Psi_3$ coupling for the 3rd generation turns out to be
of order unity, then the latter coupling will dominate
and we
will have the approximate relation $Y^b=Y^{\tau}$. For the 1st and 2nd
generations, the contributions from both couplings can be comparable. To get
something more
predictive, we may insist upon using certain texture ansatzes. For
instance, the ansatzes in \cite{Babu:1998wi} which lead to realistic fermion 
masses and mixings can be realized with the field content listed in Table
\ref{tab:422_other}, possibly supplemented by $(\bm{15},\bm{2},\bm{2})$.
[The model needs to be augmented with the Majorana coupling
$(\Phi_R^c \Psi^c)^2$ and its $C$-conjugate to give superheavy masses to the
right-handed neutrinos.] 

% If we wish to get MSSM as the low energy effective theory,
% the problem is not pairing up all the fields, which can be easily done by
% introducing a mass term between complex conjugate superfields but to pair up
% all the fields \emph{except} a pair of Higgs doublets in line with the
% missing partner mechanism which has to remain light. 

\begin{table}
\caption{The chiral superfield content of the $G_{422}$ model with two
bidoublets $H_1$, $H_2$. MSSM is recovered at low scales.}
\label{tab:422_other}
\[
\begin{array}{||c|c||c|c||}
\hline
\text{superfield} 	& \text{representation} 		&
\text{superfield} 	& \text{representation}\\
\hline
\Psi_i 			& (\bm{4},\bm{2},\bm{1}) 		&
\Psi_i^c 		& (\bm{\overline{4}},\bm{1},\bm{2})\\
\Phi_L 			& (\bm{4},\bm{2},\bm{1}) 		&
\Phi_R 			& (\bm{\overline{4}},\bm{1},\bm{2})\\
\Phi_L^c 		& (\bm{\overline{4}},\bm{2},\bm{1}) 	&
\Phi_R^c  		& (\bm{4},\bm{1},\bm{2})\\
S 			& (\bm{1},\bm{1},\bm{1})		&&\\
H_{15} 			& (\bm{15},\bm{1},\bm{1}) 		&&\\
H_1 			& (\bm{1},\bm{2},\bm{2}) 		&&\\
H_2 			& (\bm{1},\bm{2},\bm{2}) 		&&\\
\hline
\end{array}
\]
\end{table}

\begin{table}
\caption{The chiral superfield content of the $G_{422}$ model with the
$\Delta$'s and a single bidoublet.}
\label{tab:422_Delta}
\[
\begin{array}{||c|c||c|c||}
\hline
\text{superfield} 	& \text{representation} 		&
\text{superfield} 	& \text{representation}\\
\hline
\Psi_i 			& (\bm{4},\bm{2},\bm{1}) 		&
\Psi_i^c 		& (\bm{\overline{4}},\bm{1},\bm{2})\\
\Delta_L 		& (\bm{10},\bm{3},\bm{1}) 		&
\Delta_R 		& (\bm{\overline{10}},\bm{1},\bm{3})\\
\Delta_L^c 		& (\bm{\overline{10}},\bm{3},\bm{1}) 	&
\Delta_R^c  		& (\bm{10},\bm{1},\bm{3})\\
S 			& (\bm{1},\bm{1},\bm{1})		&&\\
H_{15} 			& (\bm{15},\bm{1},\bm{1}) 		&&\\
T_L 			& (\bm{1},\bm{3},\bm{1}) 		&
T_R 			& (\bm{1},\bm{1},\bm{3})\\
B 			& (\bm{1},\bm{3},\bm{3}) 		&&\\
H			& (\bm{1},\bm{2},\bm{2}) 		&&\\
\hline
\end{array}
\]
\end{table}

We would like to make a remark on the MSSM $\mu$ term. The required coupling is
$\mu H_1 H_2$, with $\mu$ on the order of the electroweak scale. 
In the present scheme, the Giudice-Masiero mechanism \cite{Giudice:1988yz} is a
plausible way to accomplish this.

Next we will look at the issue of proton decay in these $G_{422}$ models. In
order
to get proton
decay, we need $SU(3)_C$ operators like $\bm{3}\bm{3}\bm{3}$ or
$\bm{\overline{3}}\bm{\overline{3}}\bm{\overline{3}}$. This would correspond to
$SU(4)_C$ operators like $\langle \bm{4} \rangle \bm{4}\bm{4}\bm{4}$, $\langle
\bm{\overline{4}} \rangle \bm{\overline{4}}\bm{\overline{4}}\bm{\overline{4}}$,
$\bm{6} \bm{4} \bm{4}$ or $\bm{6} \bm{\overline{4}} \bm{\overline{4}}$. So far,
we have not included any couplings of this nature. (Even the
Majorana
term
$\langle \Phi_R^c \rangle
\Phi_R^c \Psi^c \Psi^c/\Lambda$ is not of this form).
In principle,
these couplings can be
forbidden by introducing a global symmetry $U(1)_{X}$ which commutes with
all the gauge symmetries. The $X$-charge assignments are given in Table
\ref{tab:X}. The proton in this case turns out to be essentially stable.

In the absence of $U(1)_X$, proton decay can occur via
dimension five operators along the lines discussed in \cite{Babu:1998wi} and
according to which, the dominant decay modes are
$\overline{\nu}K^+$ and $\overline{\nu}\pi^+$ with lifetime $\sim 10^{34-35}$
yrs.

\begin{table}
\caption{$X$-charge of the various superfields.}
\label{tab:X}
\[
\begin{array}{llllllllll}
\hline\hline
\Psi & \Psi^c & H_1 & H_2 & \Phi_L & \Phi_L^c & \Phi_R & \Phi_R^c & S & H_{15}\\
\hline
1 & -1 & 0 & 0 & 1 & -1 & -1 & 1 & 0 & 0\\
\hline\hline
\end{array}
\]
\end{table}

Next, we consider an alternative model without the missing partner
mechanism.
Recall that the missing partner mechanism is needed for two things; to give
masses to the color singlet components of $\Phi_L$ and to break the $Y^U=Y^D$
relation. The former can be taken care of by the nonrenormalizable coupling
$\Phi_L^c \Phi_L \Phi_R^c \Phi_R/\Lambda$ once the $\Phi_R$'s get a nonzero VEV.
Introducing an $SU(2)_R$ Higgs triplet
$T_R(\bm{1},\bm{1},\bm{3})$ together with its $C$-conjugate, we can arrange for
$T_R$ to get a nonzero VEV by employing the couplings $T_R^2$ and $T_R
\Phi_R^c \Phi_R$ as well as their $C$-conjugates. Without the missing partner
mechanism, we only have one bidoublet $H$ which contains both $H_u$ and $H_d$.
The up-down relation between the Yukawa couplings can be broken by the
nonrenormalizable coupling $H \langle T_R \rangle \Psi^c \Psi/\Lambda$.
However, as the cutoff scale $\Lambda$ will typically be an order of
magnitude or so larger than the $SU(2)_R$ breaking scale $M$, the splitting
between
the Yukawa couplings will only be significant for the first and second
generations and not for the third. Thus, we will still have the approximate
relation $Y^t \simeq Y^b$. 

Let us note that we may also break the $G_{422}$ symmetry with
$\Delta_L (\bm{10},\bm{3},\bm{1})$, $\Delta_R
(\bm{\overline{10}},\bm{1},\bm{3})$, $\Delta_L^c
(\bm{\overline{10}},\bm{3},\bm{1})$ and $\Delta_R^c (\bm{10},\bm{1},\bm{3})$
instead of the $\Phi$'s \cite{Mohapatra:2007af}. 
(We note that the $\mathbb{Z}_2$ matter parity of MSSM is automatically
embedded within $G_{422}$ in the absence of the $\Phi$ fields.)
The Majorana coupling will now
be
$\Delta_R^c \Psi^c \Psi^c$, and its $C$-conjugate. Under the
decomposition from $SU(4)_C$ to $SU(3)_C$, $\bm{10} \to \bm{6} \oplus
\bm{\overline{3}} \oplus \bm{1}$ and $\bm{\overline{10}} \to \bm{\overline{6}}
\oplus \bm{3} \oplus \bm{1}$. If we only have a
$(\bm{15},\bm{1},\bm{1})$ Higgs field and a singlet $S$ and we use the same
mechanism as our previous model, we find that the color sextet and triplet
components of the $\Delta$'s will get nonzero masses because the Clebsch-Gordon
contributions from $\langle S \rangle$ and $\langle H_{15} \rangle$ do not
cancel but the mass contribution to the color singlet components cancel (This
is a consequence of setting the $F$-terms to zero). This includes the
$(\bm{1},\bm{3})_{\pm 2}$
components of $\Delta_L$ and $\Delta_L^c$. Since the $\Delta_R$'s are $SU(2)_R$
triplets, the color singlet components decompose into three once $G_{422}$ is
broken. One linear combination of the $(\bm{1},\bm{1})_{0}$ components pairs up
with $S$ and the other linear combination becomes goldstone and sgoldstone
bosons and goldstinos. The $(\bm{1},\bm{1})_{\pm 2}$ components also becomes
goldstone and sgoldstone bosons and goldstinos. This leaves us with 
the $(\bm{1},\bm{1})_{\pm 4}$
components of $\Delta_R$ and $\Delta_R^c$. To realize MSSM
at low scales, we introduce the Higgs fields
$(\bm{1},\bm{3},\bm{1})$, $(\bm{1},\bm{1},\bm{3})$ and $(\bm{1},\bm{3},\bm{3})$
and the renormalizable couplings 
\begin{equation}
(\bm{1},\bm{3},\bm{1})^2, (\bm{1},\bm{1},\bm{3})^2,
(\bm{1},\bm{3},\bm{1})\Delta_L^c \Delta_L, (\bm{1},\bm{1},\bm{3})\Delta_R^c
\Delta_R, (\bm{1},\bm{3},\bm{3})^2, (\bm{1},\bm{3},\bm{3})\Delta_L \Delta_R,
(\bm{1},\bm{3},\bm{3})\Delta_L^c \Delta_R^c.
\end{equation}
The $(\bm{1},\bm{1},\bm{3})$ field acquires an
induced VEV from the $(\bm{1},\bm{1},\bm{3})\langle \Delta_R^c \rangle
\langle \Delta_R \rangle$ coupling and this pairs up the $(\bm{1},\bm{1})_{\pm
4}$ components via $\langle (\bm{1},\bm{1},\bm{3}) \rangle \Delta_R^c
\Delta_R$. The $(\bm{1},\bm{3})_{-2}$ component is paired up via
$(\bm{1},\bm{3},\bm{3})\Delta_L \langle \Delta_R \rangle$ and the
$(\bm{1},\bm{3})_2$ component is paired up via $(\bm{1},\bm{3},\bm{3})\Delta_L^c
\langle \Delta_R^c \rangle$. The $(\bm{1},\bm{3})_0$ component of
$(\bm{1},\bm{3},\bm{3})$ is self-paired (The $(\bm{1},\bm{3})_2$ and
$(\bm{1},\bm{3})_{-2}$ components of $B$ also pair up and so, what we really
have is a chain of pairings in which all the components become massive). 

The up-down Yukawa relation is broken
not by the missing partner mechanism but by the nonrenormalizable coupling
$H \langle (\bm{1},\bm{1},\bm{3}) \rangle \Psi^c \Psi/\Lambda$. (The
$C$-conjugate of this
coupling is also included.) For the same reason as before, this difference is
suppressed by $\langle (\bm{1},\bm{1},\bm{3})\rangle /\Lambda$ and so,
$Y^t \simeq Y^b$ for the third generation. Let us summarize by putting together
all the terms in the superpotential:
\begin{align}
W &\supset S(\Delta_L^c \Delta_L + \Delta_R^c \Delta_R - M^2), H_{15}(\Delta_L^c
\Delta_L + \Delta_R^c \Delta_R), H_{15}^2,\nonumber\\
&\quad H\Psi^c \Psi, (H T_L
\Psi^c\Psi + H T_R \Psi^c\Psi)/\Lambda, H H_{15}
\Psi^c\Psi/\Lambda,\Delta_L^c \Psi \Psi + \Delta_R^c \Psi^c \Psi^c\nonumber\\
&\quad T_L^2 + T_R^2, T_L \Delta_L^c \Delta_L + T_R
\Delta_R^c \Delta_R, B^2, B\Delta_L \Delta_R + B \Delta_L^c \Delta_R^c
\end{align}

One consequence of $C$-parity is the existence of $\mathbb{Z}_2$ domain walls
once it is spontaneously broken. The scale at which this occurs happens to
be same
as the $G_{422}$ breaking scale. Such domain walls can give rise to cosmological
problems unless they are inflated away. In addition,
when $G_{422}$ breaks down to MSSM, magnetic monopoles carrying two quanta of
Dirac magnetic charge \cite{Lazarides:1980cc} can be generated. Astrophysical
and cosmological bounds on such
monopoles are fairly stringent and the standard solution is to inflate
them away. One way to do this in our case is to invoke shifted hybrid inflation
\cite{Jeannerot:2000sv} where
a nonrenormalizable term
($S\left[(\Phi_L^c\Phi_L)^2+(\Phi_R^c\Phi_R)^2\right]$)
is added to the superpotential. With such a
suitably altered model, it is possible to start inflation with a 
trajectory where $C$-parity as well as $G_{422}$ are already spontaneously
broken. In such a scenario, domains do not form once
inflation ends and neither do monopoles. During inflation
itself, the inflationary trajectory is identical to that analyzed in
\cite{Jeannerot:2000sv} with $\Phi_L=\Phi_L^c=0$ throughout.
Although the postinflationary trajectory will be
different, the end result is that both $C$-domain walls and monopoles are
eliminated.

% At the end of inflation, an instability in the $\Phi$ directions will appear
% and the universe will slide down and slowly settle down to the minimum of the
% potential after quite a number of oscillations. This is due to the presence of
% friction, which is caused, among many other things by the Majorana generating
% seesaw term $(\Phi_R^c \Psi^c)^2/\Lambda$
% \cite{Pati:2003qia,Lazarides:1996dv}.
% The instability in the $\Phi$ directions will lead to a large nonzero value
% for
% $\Phi_R^c$ and this will lead to the pair production of right-handed
% neutrinos.
% In the process of decaying, provided that we have some CP-violation, these
% right-handed neutrinos will decay leaving us will a slight excess of leptons
% over antileptons. Weak sphalerons will then convert some of these excesses
% into
% an excess of quarks over antiquarks, leading to the observed asymmetry between
% matter and antimatter that we observe today. This process is known as
% baryogenesis via leptogenesis \cite{Lazarides:1996dv}. 

The end of inflation is followed by the decay of the inflaton fields $\Phi_R$,
$\Phi_R^c$ and $S$ into right handed neutrinos ($\nu^c$) and sneutrinos
($\widetilde{\nu^c}$). As discussed in \cite{Lazarides:1996dv,Pati:2003qia},
following \cite{Fukugita:1986hr}, the subsequent out of equilibrium decay of
$\nu^c$ and $\widetilde{\nu^c}$ generates lepton asymmetry, which is then
partially converted to the observed baryon asymmetry by the electroweak
sphalerons.

Note that hybrid inflation requires that the $G_{422}$ breaking scale is
comparable
to $M_{\text{GUT}}\sim 10^{16}$ GeV. If $G_{422}$ breaks at significantly lower
(such as intermediate)
scales, an alternative scenario for suppressing monopoles should be employed
(see, for instance, \cite{Dar:2006cm}).

% Supersymmetric $G_{422}$ models with $C$-parity which reduces to MSSM at low
% energy
% scales are possible and we have shown how to construct some such minimal
% models. We will list some of the major differences from prior models which
% have
% been studied in the literature:
% \begin{itemize}
%  \item We use $(H_{15},\bm{1},\bm{1})$ Higgs superfields instead of
% $(\bm{6},\bm{1},\bm{1})$ superfields to pair up the additional states.
%  \item Unlike Ref. \cite{Melfo:2003xi}, we do not resort to any fine-tuning to
%  get MSSM.
%  \item We invoke $C$-parity.
% \end{itemize}

In conclusion, we have constructed realistic supersymmetric $G_{422}$ models in
which the scalar (Higgs) sector respects a 
discrete left-right symmetry ($C$-parity) which is spontaneously broken at 
the same scale as the $SU(2)_R$ gauge symmetry. We have shown how the MSSM 
is recovered without fine tuning at low scales. The scalar fields we 
employ enable us to reproduce, following \cite{Babu:1998wi}, the observed
fermion 
masses and mixings, implement a missing partner mechanism \cite{Lee:2007aa}, and
realize 
inflation followed by non-thermal leptogenesis.

\begin{acknowledgments}
This work is supported by DOE Grant No. DE-FG02-91ER40626.
\end{acknowledgments}

\end{document}